\documentclass[runningheads]{llncs}
\usepackage[T1]{fontenc}
\usepackage{type1cm}        
\usepackage{makeidx}         
\usepackage{graphicx}        
\usepackage{multicol}        
\usepackage[bottom]{footmisc}

\usepackage{newtxtext}       %
\usepackage[varvw]{newtxmath}       
\usepackage{mathptmx}       
\usepackage{helvet}         
\usepackage{courier}        
\usepackage{type1cm}        

\usepackage{makeidx}         
\usepackage{graphicx}        
\usepackage{multicol}        
\usepackage[bottom]{footmisc}
 \usepackage{subcaption}
\usepackage{graphicx}
\begin{document}
\title{A Novel Feature Extraction Model for the Detection of Plant Disease from Leaf Images in Low Computational Devices }
%
%

\author{Rikathi Pal \inst{1}\orcidID{0009-0002-8971-1820}\\ Anik Basu Bhaumik \inst{1}\orcidID{0009-0000-0778-2234} \\ Arpan Murmu \inst{1}\orcidID{0000-0003-1378-4484}\\ Sanoar Hossain \inst{1}\orcidID{0000-0002-1232-7487} \\ Biswajit Maity\inst{2}\\ Soumya Sen \inst{1}\orcidID{0000-0002-9178-6410}}
\authorrunning{Rikathi et al.}
%
\institute{A. K. Choudhury School of Information Technology, University of
Calcutta, Kolkata
\email{rikathi.pal@gmail.com}\\\email{anikbb@gmail.com}\\\email{murmuarpan530@gmail.com}\\\email{snr.hossain12@gmail.com}\\\email{iamsoumyasen@gmail.com}
\and
Department of Computer Application and Science, Institute of Engineering and Management, Kolkata\\
\email{biswajit.maity1@gmail.com}}
\maketitle              
\begin{abstract}
Diseases in plants cause significant danger to productive and secure agriculture. Plant diseases can be detected early and accurately, reducing crop losses and pesticide use. Traditional methods of plant disease identification, on the other hand, are generally time-consuming and require professional expertise. It would be beneficiary to the farmers if they could detect the disease quickly by taking images of the leaf directly. This will be a time-saving process and they can take remedial actions immediately. To achieve this a novel feature extraction approach for detecting tomato plant illnesses from leaf photos using low-cost computing systems such as mobile phones is proposed in this study. The proposed approach integrates various types of Deep Learning techniques to extract robust and discriminative features from leaf images. After the proposed feature extraction comparisons has been done on five cutting-edge deep learning models: AlexNet, ResNet50, VGG16, VGG19, and MobileNet. The dataset contains 10,000 leaf photos from ten classes of tomato illnesses and one class of healthy leaves. Experimental findings demonstrate that AlexNet with an accuracy score of 87\%, with the benefit of being quick and lightweight, making it appropriate for use on embedded systems and other low-processing devices like smartphones.

\end{abstract}

\section{Introduction}
\label{intro}
\begin{figure}
    \centering
    \includegraphics[width= \textwidth]{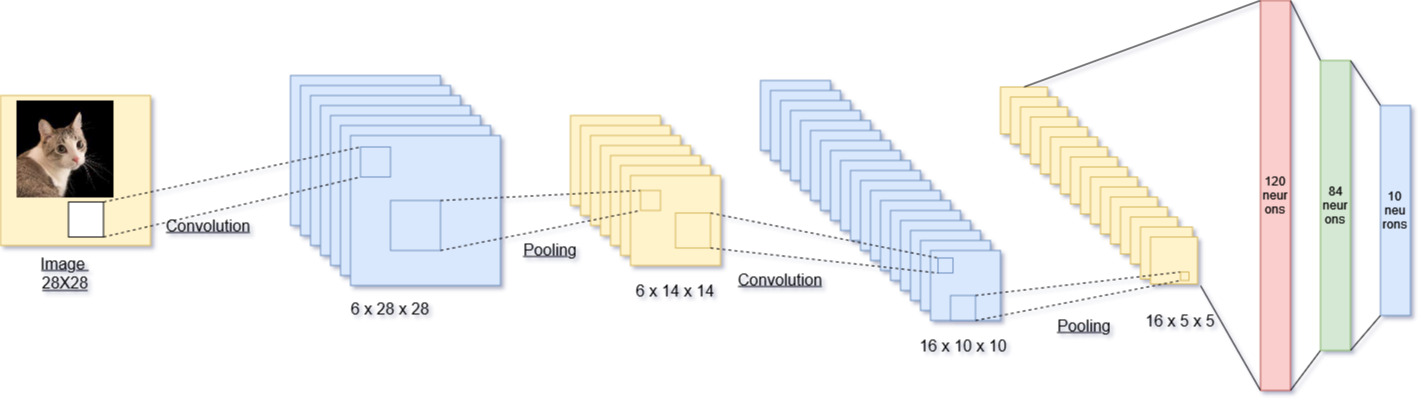}
    \caption{The Le-Net CNN model }
    \label{fig:conv}
\end{figure}

India's economy has mostly been driven by agriculture, hence research is essential in this field to improve productivity and food quality keeping profit in mind. 
These diseases are brought on by pathogens like fungi, bacteria, viruses and other environmental factors. Diseases in the tomato plant are typically brought on by bacteria and fungi and it is crucial to diagnose a disease in its early stages \cite{kulkarni2012applying}. It is not possible for the farmer on the field to examine his specimens for detecting diseases every time. So, agricultural production is continuously monitored, but doing so could be prohibitively expensive and time-consuming \cite{argenti1990fast}. Hence there is a need for a more practical and less expensive way for real-time detection of leaf diseases.
Due to the current wave in technology in India about 70\% of the rural dwellers in most states possess a smartphone. This gives computer vision and deep learning technicians a chance to solve this problem, offering an optimized image-based process control and autonomous inspection of diseases. This paper aims for a refined approach through which the farmers could run an application on their low-computation devices and still receive effective results. Efforts have been made towards feature extraction to reduce the number of features for faster computation and then in the selection of a model that gives better results. 
Neural networks are algorithms that mimic the learning process of the human brain by efficiently extracting features from data. They consist of interconnected layers of neurons with adjustable weights. Through iterative training and backpropagation, neural networks uncover patterns and delicate nuances in data. These properties are transforming various industries such as computer vision, natural language processing, healthcare, and robotics \cite{muller1995neural}. Convolution neural networks (CNN), a feed-forward model are among the most important neural networks in the deep learning industry designed for processing organized arrays of data and it has received a lot of interest from both industry and academia for recognizing design in the input image. One of the initial models that proved the effectiveness of CNN is the Lenet Model\cite{726791}  as shown in fig \ref{fig:conv}.
CNNs use convolution layers \cite{li2021survey} to extract features in image processing. These layers use small filters that slide across input images, enabling automatic recognition of edges, textures, and patterns. CNNs learn hierarchical representations of features by stacking multiple convolution layers and use shared weights for efficiency and to reduce overfitting \cite{Liu_2015_CVPR}. CNNs revolutionize image tasks with superior spatial detection, powering AI vision for intricate pattern recognition.

The goal is to identify plant diseases using just the sections in the leaves that show symptoms of the diseases. Instead of the whole leaf image, it is more beneficial to extract the essential parts such that the CNN becomes aware of the nuances of the disease and changes its kernel weights in a more detailed fashion. A CNN works better when fueled with data and a good amount of computational resources, due to the presence of limited resources and data by trying to model the dataset by highlighting only essential features for an efficient feature space formulation in the model designed. 
The remainder of this paper is structured as follows: In Section \ref{relworks}, the paper delves into a comprehensive discussion of the related work and the various positions this study holds within the existing literature. Section \ref{meth} presents a detailed outline of the methodology that has been put forth for this research endeavor. Moving forward, Section \ref{results} offers a comprehensive presentation of the experimental results obtained. Lastly, in Section \ref{conc}, concluding remarks are offered that summarize the key findings of our study and shed light on the potential future directions for our research.

\section{Related Works}\label{relworks}

Classifying diseases using leaf images is a challenging task. Various deep-learning methods have been proposed for this purpose. In this section, existing techniques are reviewed and our approach for disease classification in tomato leaves in low computational devices.
In paper \cite{sardogan2018plant}, the authors developed a CNN model based on RGB components of tomato leaf images from the plantvillage dataset, where the learning vector quantization (LVQ) algorithm is used as a classifier. LVQ is a neural network algorithm that combines competitive and supervised learning, making it suitable for classification tasks. In paper \cite{rastogi2015leaf}, the proposed system encompasses two vital phases for comprehensive plant health assessment. In the initial phase, plant recognition relies on leaf feature analysis, involving image pre-processing and feature extraction. An Artificial Neural Network (ANN) is then employed for classification. which focuses on disease classification, entailing K-means-based segmentation of diseased areas, feature extraction, and ANN-based disease categorization.
Paper \cite{shoaib2023advanced} reviews recent research on using deep learning, including CNNs, RNNs, GANs, and Transformers, for plant disease identification. They address challenges like data availability, image quality, and distinguishing healthy and diseased plants. The study offers valuable insights, summaries of recent research, and recommendations for overcoming these challenges, assessing current approaches, and suggesting new directions.
In paper \cite{sakkarvarthi2022detection} they employed a CNN-based approach for classifying tomato leaf diseases, using a dataset of 10k images with 10 different diseases. They enriched the dataset with data augmentation techniques like rotations, flips, and scaling. Their CNN model outperformed three established pre-trained models, InceptionV3, ResNet 152, and VGG19.
As recent studies have advanced tomato disease classification with deep learning, further research is needed as there is a need to explore diseases affecting leaves. Larger and more diverse datasets are crucial for robust models. Model interpretability is vital for user confidence, especially among farmers and specialists. Understanding the impact of data augmentation, especially for imbalanced datasets, is essential for enhancing tomato disease classification and agricultural disease management
In conclusion, recent research has significantly enhanced tomato disease classification using deep learning, particularly CNNs. Various approaches, including Fuzzy-SVM and hybrid models, have shown promise. However, gaps remain, such as the need for larger datasets, improved interpretability, and expanding disease classification to stems and fruits. Addressing these gaps will support sustainable tomato farming practices.

\section{Proposed methodology}\label{meth}
\subsection{Data Pre-processing}
\begin{figure}[h]
    \centering
    \begin{subfigure}{0.45\textwidth}
        \includegraphics[width=\linewidth, height=1.5in]{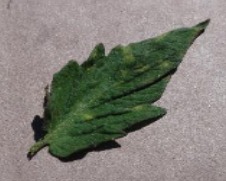}
        \label{fig:initial1}
    \end{subfigure}
    \hfill
    \begin{subfigure}{0.45\textwidth}
        \includegraphics[width=\linewidth, height=1.5in]{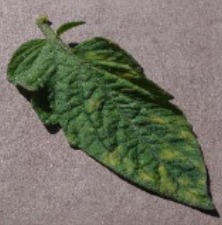}
        \label{fig:initial2}
    \end{subfigure}    
       \caption{Image before preprocessing}
       \label{fig:initial}
\end{figure}

\begin{figure}[h]
    \centering
    \begin{subfigure}{0.45\textwidth}
        \includegraphics[width=\linewidth, height=1.5in]{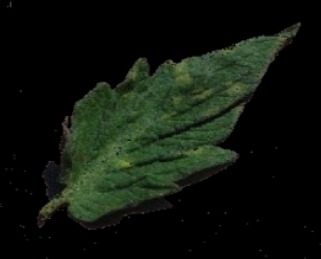}
    \end{subfigure}
    \hfill
    \begin{subfigure}{0.45\textwidth}
        \includegraphics[width=\linewidth, height=1.5in]{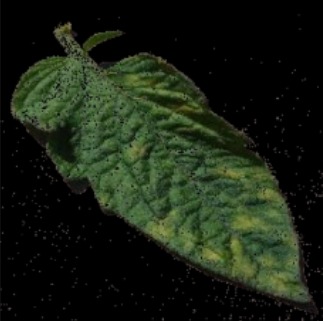}
    \end{subfigure}

    \caption{Image with background eliminated}
    \label{fig:final}
\end{figure}

The dataset consisted of each image of the form as shown in Fig \ref{fig:initial}. The background contains unnecessary features, hence it is important to focus only on the important features of the leaf. The New Plant Diseases Dataset  \footnote{https://www.kaggle.com/datasets/vipoooool/new-plant-diseases-dataset/data/} is used, from where the tomato leaf disease dataset is collected. This is an augmented dataset, where each image is of size 256 X 256 X 3. The dataset comprises of close to 10000 RGB images of crop leaves in both healthy and diseased conditions, divided into 10 different classes. The total dataset dedicates 80\% of its data to training and the rest to validation. As a first step in preprocessing only the leaf part was extracted from the image. It was done using the pixel filter mechanism,  the background of the image was filtered out and assigned the pixel value [0,0,0] which is black color as seen in Fig \ref{fig:final}.

\begin{figure}[h]
    \centering
    \begin{subfigure}{0.45\textwidth}
        \includegraphics[width=\linewidth, height=1.5in]{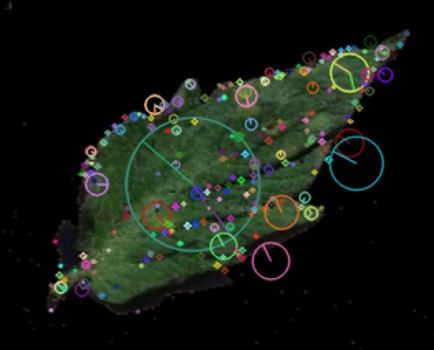}
        \label{fig:final2}
    \end{subfigure}
    \hfill
    \begin{subfigure}{0.45\textwidth}
        \includegraphics[width=\linewidth, height=1.5in]{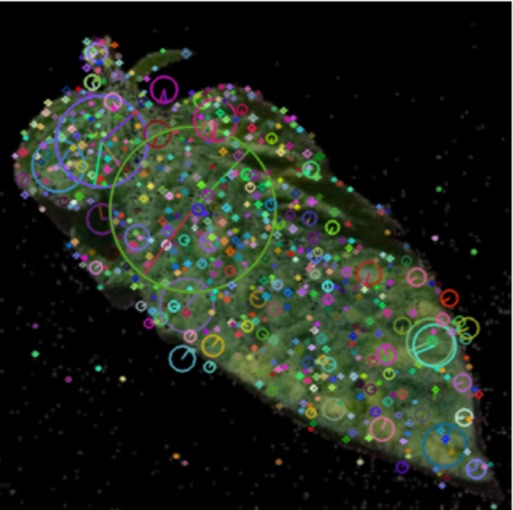}
        \label{fig:final2}
    \end{subfigure}

    \caption{Training key points of the image after applying SIFT}
    \label{fig:SIFT}
\end{figure}

Then the SIFT (Scale Invariant Feature Transform) operator was used to detect and describe the local features of the image \cite{lindeberg2012scale}. SIFT is a powerful feature extraction technique used extensively in computer vision problems for identifying and describing distinctive features in images. It achieves scale, rotation, and illumination invariance by detecting the local extrema in the image's scale space, refining key locations, assigning orientations, and generating unique descriptors based on local gradient information. These key points and descriptors in SIFT enable robust image matching, object recognition, and tracking.
In Fig \ref{fig:SIFT} the circles of various sizes throughout the leaf which indicates the training keypoints within the leaf along with the size of each keypoint.
\begin{figure}[h]
    \centering
    \includegraphics[width=\linewidth]{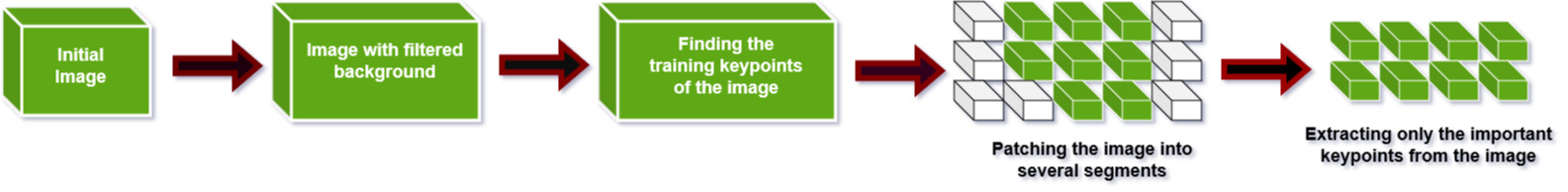}
    \caption{Extraction of important key points from image }
    \label{fig:patch}
\end{figure}
Then the image is broken down into several small patches focusing on the training key points that is obtained by applying SIFT, therefore the patches will contain only the important features of the leaf that is the infected portion of the leaf. This pre-processing of the image will reduce the image size by keeping only the main features required for image processing and to detect the leaf diseases.
The flow of the entire work is mentioned in fig \ref{fig:patch}

\subsection{Model Selection}\label{modsel}
The method implied here utilizes the features learned by previously developed models and uses them for a customized task through fine-tuning. Feature representations of different models have been used in the experiment through the combined use of those models. This work utilizes several large image models like ResNet-50, VGG-16,  VGG-19, AlexNet, MobileNet, and a custom model combining convolutional neural networks and linear binary patterns (CNN-LBP). 

\subsubsection{The VGG Networks:}

The VGG architectures \cite{simonyan2015deep} are known for their simplicity and efficiency in image classification tasks. The VGG architectures are deep network architectures with 16 or 19-layer variants, giving them the capacity to learn more features from images. These networks follow uniformity in design with smaller (3x3) kernels, stacked on top of one another succeeded by a pooling layer, and finally for inference dense layers of size 4096 are present. An increase in smaller size filters enables the capturing of a rich set of image features more effectively than before. The VGG-16 and VGG-19 architectures and trained using the data with a validation accuracy of about 94\%. This is a testament to the strength of simplicity and depth in VGG Networks. ImageNet weight for these models are initially used and then the model is fine-tuned for the particular task. 

\subsubsection{The ResNet Architecture:}

The ResNet architecture \cite{he2016deep} or residual network architecture was developed to develop a learning framework that was easy to train but also deep enough networks to compete with other large networks. The layers are used to learn residual functions concerning layer inputs. With the addition of multiple layers to a network comes the problem of diminishing and amplifying gradients, this can be solved by inter-layer normalization however, the problem of degradation involves saturation of weights, and residual mapping as formulated by He et al, is an efficient solution to the problem. In the case of a normal learning process in networks, it tries to learn from a normal mapping say H(x). However, with residual mapping, the stacked layers are made to fit with a mapping of F(x)=H(x) - x which is the difference between the desired output and the output of the layer. It was hypothesized with results that this residual learning method yielded results with a faster training process. The pre-trained ImageNet weight for these models are used initially and then the model is fine-tuned for our particular task. 

\subsubsection{AlexNet:} 
AlexNet is a deep convolutional network developed by Alex Krizhevzky \cite{NIPS2012_c399862d} for the ILSVRC challenge in 2012. It consists of 8 layers of learnable parameters, it has 5 convolution layers followed by three fully connected layers with a consecutive Rectified linear unit (RELU) activation for the hidden layers. The training was done for the network using about 1.2 million images. The training for this network was done on 2 GPUs. This was a model developed in the initial stages of the deep learning scenario. This demonstrated what a large network could do when fueled with a huge number of data points. The model has been developed from scratch and trained using our dataset.

\subsubsection{CNN-LBP Model:}
The CNN-LBP model uses local binary convolution\cite{juefei2017local}(LBC). This was a model that was developed in an attempt to reduce the complexity of CNNs. The Local binary convolution layer is used to estimate the output from the non-linear activations of the Convolution layers. The LBC layer consists of sparse matrix binary filters, a non-linear activation function, and linear weights which contain the weighted average of the convolutional response maps. These models were tested for lower tendencies to overfit and for inference scenarios of CNNs.

\subsubsection{MobileNet:}
MobileNet is a compact and efficient deep learning neural network architecture designed specifically for mobile and embedded devices \cite{howard2017mobilenets}. It utilizes depthwise separable convolutions which reduces computational complexity while maintaining high accuracy in tasks like image classification and object detection. MobileNet models are known for their small memory footprint and fast inference speed, making them ideal for applications where resource constraints are a concern, such as mobile apps, robotics, and IoT devices.

\section{Results}\label{results}
The models mentioned above were trained using the processed dataset. The models yielded results as shown below. 

\begin{figure}
    \centering
    \resizebox{0.9\textwidth}{!}{%
        \begin{subfigure}{0.32\textwidth}
            \includegraphics[width=\linewidth]{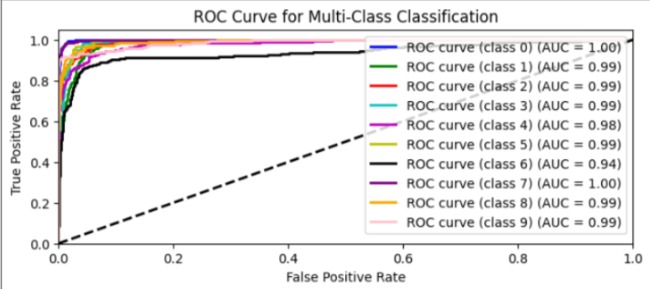}
            \caption{Alexnet}
            \label{fig:initial1}
        \end{subfigure}
        \hfill
        \begin{subfigure}{0.32\textwidth}
            \includegraphics[width=\linewidth]{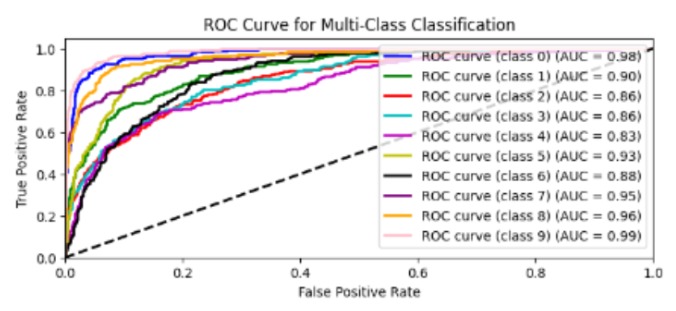}
            \caption{Resnet}
            \label{fig:initial2}
        \end{subfigure}
        \hfill
        \begin{subfigure}{0.32\textwidth}
            \includegraphics[width=\linewidth]{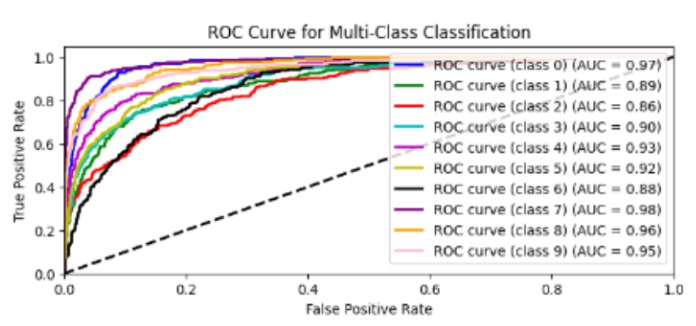}
            \caption{VGG16}
            \label{fig:initial3}
        \end{subfigure}
    }
    
    \medskip
    
    \resizebox{0.9\textwidth}{!}{%
        \begin{subfigure}{0.32\textwidth}
            \includegraphics[width=\linewidth]{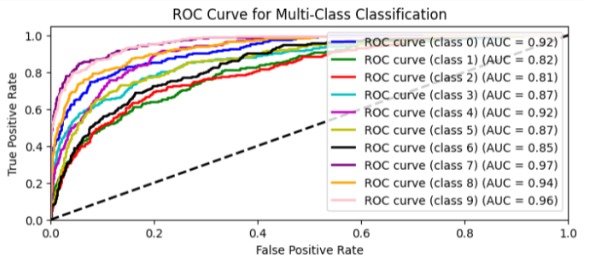}
            \caption{VGG19}
            \label{fig:initial4}
        \end{subfigure}
        \hfill
        \begin{subfigure}{0.32\textwidth}
            \includegraphics[width=\linewidth]{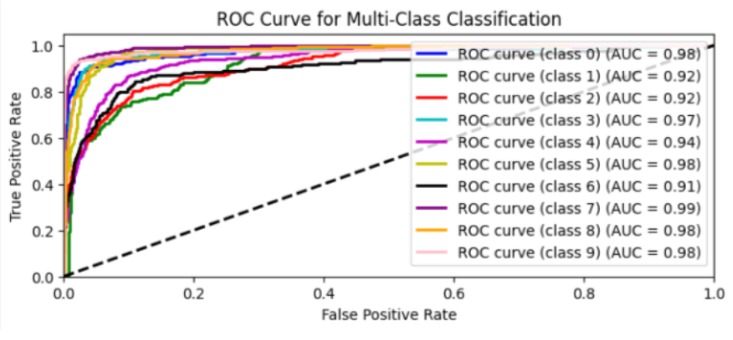}
            \caption{CNN + LBP}
            \label{fig:initial5}
        \end{subfigure}
        \hfill
        \begin{subfigure}{0.32\textwidth}
            \includegraphics[width=\linewidth]{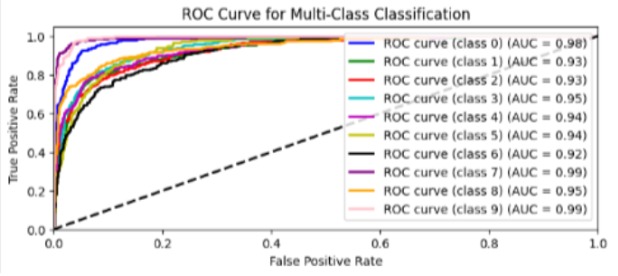}
            \caption{MobileNet}
            \label{fig:initial6}
        \end{subfigure}
    }
    
    \caption{Receiver Operating Characteristic Curves (ROC) of the tried models}
    \label{fig:ROC}
\end{figure}

\begin{figure}
    \centering
    \resizebox{0.8\textwidth}{!}{%
        \begin{subfigure}{0.32\textwidth}
            \includegraphics[width=\linewidth]{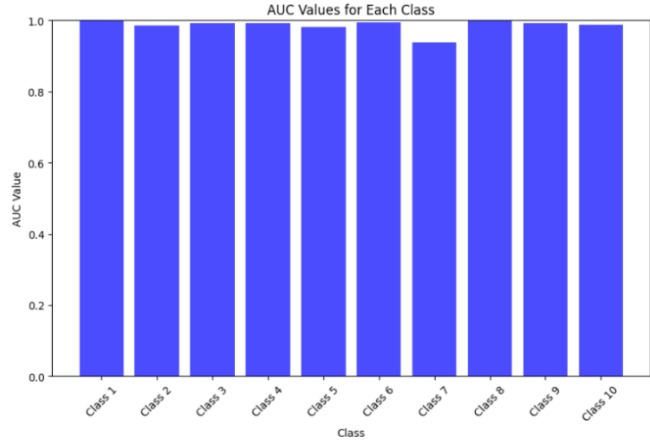}
            \caption{Alexnet}
            \label{fig:initial1}
        \end{subfigure}
        \hfill
        \begin{subfigure}{0.32\textwidth}
            \includegraphics[width=\linewidth]{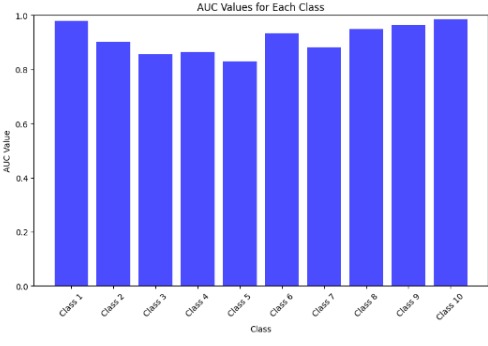}
            \caption{Resnet}
            \label{fig:initial2}
        \end{subfigure}
        \hfill
        \begin{subfigure}{0.32\textwidth}
            \includegraphics[width=\linewidth]{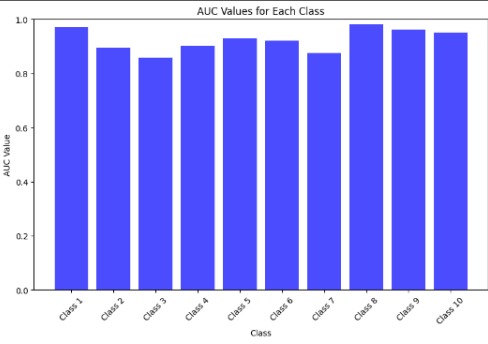}
            \caption{VGG16}
            \label{fig:initial3}
        \end{subfigure}
    }
    
    \medskip
    
    \resizebox{0.8\textwidth}{!}{%
        \begin{subfigure}{0.32\textwidth}
            \includegraphics[width=\linewidth]{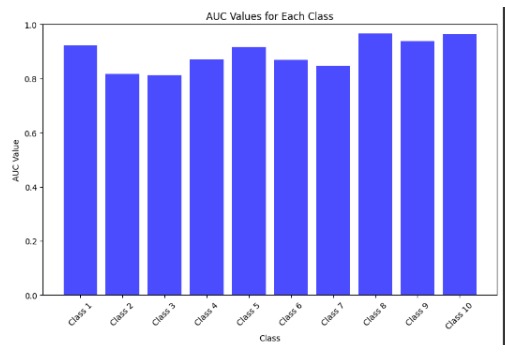}
            \caption{VGG19}
            \label{fig:initial4}
        \end{subfigure}
        \hfill
        \begin{subfigure}{0.32\textwidth}
            \includegraphics[width=\linewidth]{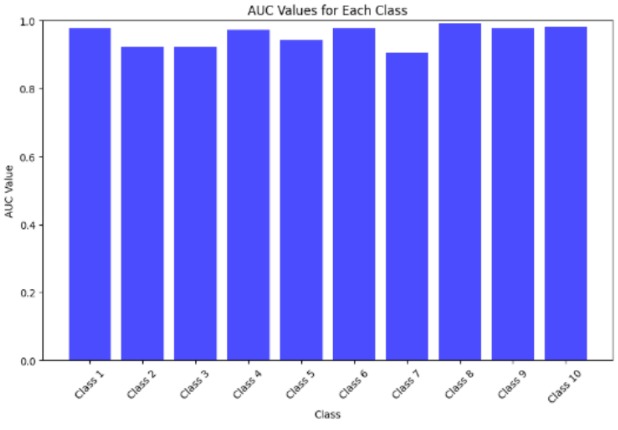}
            \caption{CNN + LBP}
            \label{fig:initial5}
        \end{subfigure}
        \hfill
        \begin{subfigure}{0.32\textwidth}
            \includegraphics[width=\linewidth]{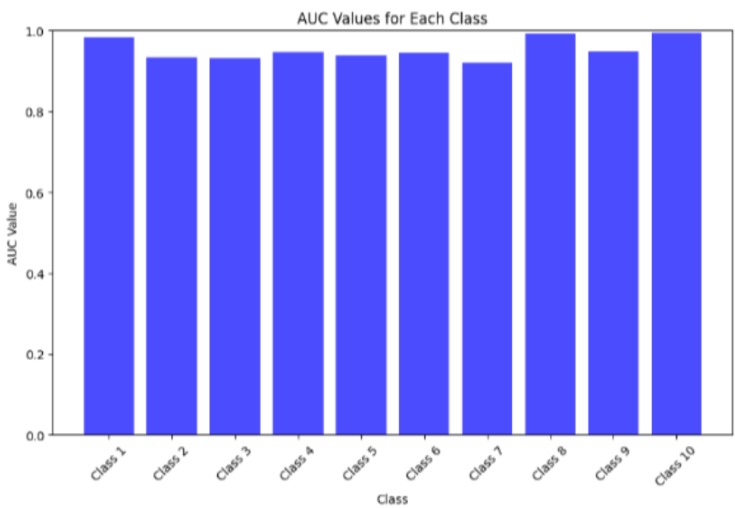}
            \caption{MobileNet}
            \label{fig:initial6}
        \end{subfigure}
    }
    
    \caption{Area under the Curves (AUC) of the tried models}
    \label{fig:AUC}
\end{figure}

\begin{table}[htbp]
    \centering
    \begin{tabular}{|l|c|c|c|c|}
        \hline
        \textbf{Model} & \textbf{Accuracy} & \textbf{Precision} & \textbf{Recall} & \textbf{F1 Score} \\
        \hline
        AlexNet & \textbf{87\%} & \textbf{0.86} & \textbf{0.85} & \textbf{0.87} \\
        ResNet50 & 54\% & 0.60 & 0.54 & 0.52 \\
        Vgg16 & 58\% & 0.66 & 0.56 & 0.57 \\
        Vgg19 & 56\% & 0.65 & 0.59 & 0.59 \\
        CNN-LBP & 79\% & 0.80 & 0.78 & 0.79 \\
        MobileNet & 67\% & 0.73 & 0.65 & 0.65 \\
        \hline
    \end{tabular}
    \caption{Model Metrics}
    \label{tab:model_metrics}
\end{table}
The fig \ref{fig:confusion} below shows the confusion matrix of all the models that are tried.  The ROC (Receiver Operating Characteristic) curve in fig: \ref{fig:ROC} and AUC (Area Under the Curve) in fig: \ref{fig:AUC} are used for the performance classification of the machine learning models. The ROC curve that the classifiers are giving curves closer to the top-left corner which clearly indicates that Alexnet is giving the best accuracy when it deals with just the important features of the image. Again, while looking into the AUC curve of the image can clearly see that the area under the curve is maximum which also proves that the Alexnet is giving the best result.

\begin{figure}
\centering
\resizebox{0.8\textwidth}{!}{%
    
    \begin{subfigure}{0.32\textwidth}
        \includegraphics[width=\linewidth]{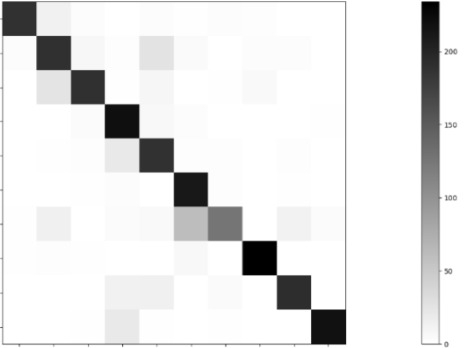}
        \caption{Alexnet}
        \label{fig:initial1}
    \end{subfigure}
    \hfill
    \begin{subfigure}{0.32\textwidth}
        \includegraphics[width=\linewidth]{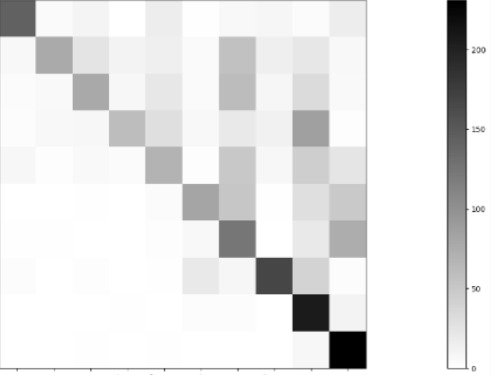}
        \caption{Resnet}
        \label{fig:initial2}
    \end{subfigure}    
    \hfill
    \begin{subfigure}{0.32\textwidth}
        \includegraphics[width=\linewidth]{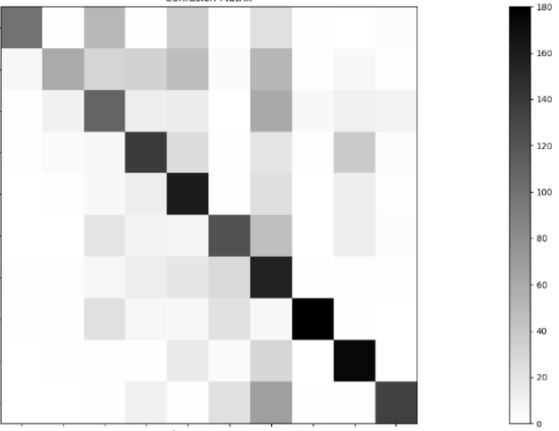}
        \caption{VGG16}
        \label{fig:initial3}
    \end{subfigure}
    }
    \medskip
    \resizebox{0.8\textwidth}{!}{%

    \begin{subfigure}{0.32\textwidth}
        \includegraphics[width=\linewidth]{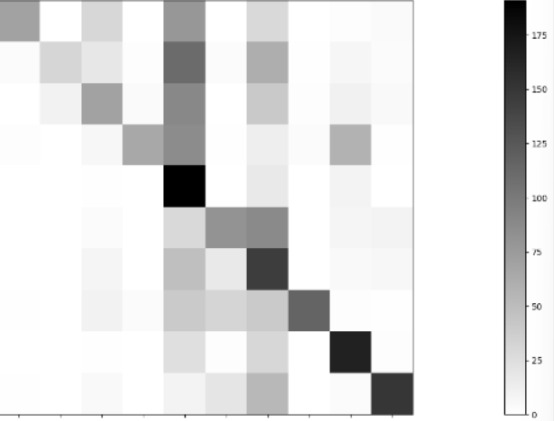}
        \caption{VGG19}
        \label{fig:initial4}
    \end{subfigure}
    \hfill
    \begin{subfigure}{0.32\textwidth}
        \includegraphics[width=\linewidth]{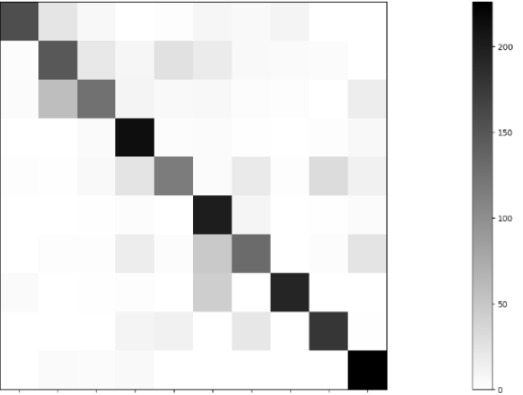}
        \caption{CNN + LBP}
        \label{fig:initial5}
    \end{subfigure}    
    \hfill
    \begin{subfigure}{0.32\textwidth}
        \includegraphics[width=\linewidth]{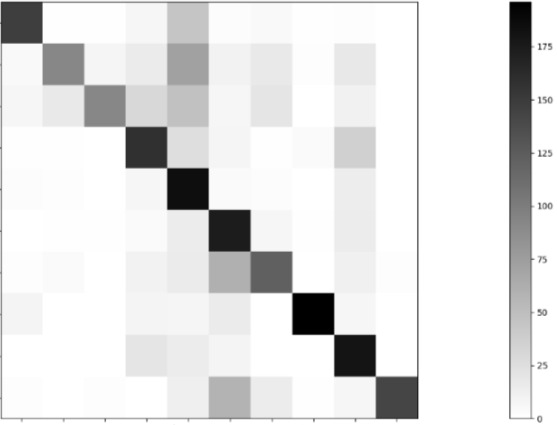}
        \caption{MobileNet}
        \label{fig:initial6}
    \end{subfigure}
    }
    \caption{Confusion matrices from the models}
    \label{fig:confusion}
    
\end{figure}

The above results manifest that AlexNet supersedes other deep learning models while working on the preprocessed images that contain only the important features of the image. Hence in practical applications where the farmers are going to use an app to detect plant diseases, the app will be built based on AlexNet to provide them most accurate results in low computational devices. 

\section{Conclusion}\label{conc}
The extensive findings and analysis of the images infer that the proposed image pre-processing technique excels at extracting the features from images, leveraging the powerful SIFT operator, and then generating patches that encapsulate the most pertinent information. The results obtained through experimentation demonstrate the remarkable effectiveness of this methodology particularly when training compact networks like AlexNet. It is going to work fine on the low computational resources. This is apt for the farmer while they will use it to detect plant diseases in real-time on devices like a smart phone.  
As an additional attempt, model compression, and optimization approaches can be applied to these models to produce a more resource-friendly and computationally simple solution to the problem of inference and training at the edge.

\bibliographystyle{splncs04}
\bibliography{reference.bib}

\end{document}